\documentclass[11pt]{article}
\usepackage{mathptmx}
\usepackage{sprinconf}
\usepackage{bm}
\usepackage{amsmath,amsfonts,amssymb,amsthm}
\def \<{\langle} 
\def \>{\rangle}

\def\tt{\otimes}                               
\def\bt{\boxtimes}
\def\<{\langle}
\def\>{\rangle}

\def\d{\partial}

\def\st{\; | \;}                               

\def\z{{\mathrm{z}}}
\def\w{{\mathrm{w}}}
\def\u{{\mathrm{u}}}

\def\bt{{\mathrm{T}}}
\def\bla{\lambda}
\def\bchi{\chi}
\def\sip{{\mathrm{s.p.}}}

\newcommand{\CC}{\mathbb{C}}       
\newcommand{\RR}{\mathbb{R}}       
\newcommand{\NN}{\mathbb{N}}       
\newcommand{\ZZ}{\mathbb{Z}}       

\def\al{\alpha}                         
\def\be{\beta}

\def\de{\delta}

\def\la{\lambda}

\def\si{\sigma}

\def\g{{\mathfrak{g}}}      
\def\h{{\mathfrak{h}}}      

\def\so{{\mathfrak{so}}}

\def\sp{{\mathfrak{sp}}}

\def\T{\mathrm{T}}           
\DeclareMathOperator{\Cur}{Cur}
\DeclareMathOperator{\Der}{Der}
\DeclareMathOperator{\Res}{Res}

\DeclareMathOperator{\End}{End}

\begin{document}

\title{Vertex (Lie) algebras in higher dimensions}


\author{Bojko Bakalov}

\affil{Department of Mathematics, North Carolina State University,
Raleigh, NC 27695, USA}

\vspace{-24pt}\noindent
\emph{Hermann Weyl prize lecture presented at the 
26th International Colloquium on Group Theoretical Methods 
in Physics, New York, June 27, 2006}
\vspace{12pt}

\beginabstract
Vertex algebras provide an axiomatic algebraic description of the operator
product expansion (OPE) of chiral fields in 2-dimensional conformal field
theory. Vertex Lie algebras (= Lie conformal algebras) encode the singular part 
of the OPE, or, equivalently, the commutators of chiral fields. 
We discuss generalizations of vertex algebras and vertex Lie algebras,
which are relevant for higher-dimensional quantum field theory.
\endabstract

\vspace{-12pt}

\section{Vertex algebras and Lie conformal algebras}\label{s1} 

In the theory of \emph{vertex algebras} \cite{ref1,ref2,ref3},
the (quantum) \emph{fields} are linear maps
from $V$ to $V[\![z]\!] [z^{-1}]$, where $z$ is a formal variable.
They can be viewed as formal series
$a(z) = \sum_{n\in\ZZ} \, a_{(n)} \, z^{-n-1}$ with
$a_{(n)} \in \End V$ such that $a_{(n)} b = 0$ for $n$ large enough.
Let $\Res \, a(z) = a_{(0)}$; then the \emph{modes} of $a(z)$ 
are given by $a_{(n)} = \Res \, z^n a(z)$.
The \emph{locality} condition for two fields
\begin{equation*}
(z-w)^{N_{ab}} \, [a(z), b(w)] = 0, 
\qquad N_{ab} \in \NN
\end{equation*}
is equivalent to the \emph{commutator formula} 
\begin{equation*}
[a(z), b(w)] = \sum_{j=0}^{N_{ab}-1} c_j(w) \, \d_w^j \de(z-w) / j!
\end{equation*}
for some new fields $c_j(w)$, where
$\de(z-w)$ is the formal delta-function (see \cite{ref2}).
The \emph{operator product expansion} (OPE) can be written symbolically
\begin{equation*}
a(z) b(w) = \sum_{j\in\ZZ} \, c_j(w) \, (z-w)^{-j-1}
\end{equation*}
(see \cite{ref2} for a rigorous treatment).
The new field $c_j$ is called the \emph{$j$-th product} of $a,b$ and
is denoted $a_{(j)} b$. The \emph{Wick product} (= normally ordered product) 
coincides with $a_{(-1)} b$.
The $j$-th products satisfy the \emph{Borcherds identity}
\cite[Eq (4.8.3)]{ref2}.
Finally, recall that every vertex algebra $V$ is endowed with a 
\emph{translation operator\/} $T \in \End V$ satisfying
$[T, a(z)] = \d_z \, a(z) = (Ta)(z)$.

The commutator of two fields is encoded by the singular part of their
OPE and is uniquely determined by their $j$-th products for $j\ge0$.
The \emph{$\la$-bracket}
\begin{equation*}
[a_\la b] = \Res_z \, e^{z\la} a(z)b
= \sum_{j=0}^{N_{ab}-1} \la^{\! j} \, a_{(j)} b / j!
\end{equation*}
satisfies the axioms of a \emph{Lie conformal algebra} introduced by Kac
\cite{ref2} (also known as a \emph{vertex Lie algebra}; cf.\ \cite{ref3,ref4}).
This is a $\CC[T]$-module $R$
with a $\CC$-linear map $R\tt R \to R[\la]$ satisfying:  

\smallskip
\emph{sesquilinearity}  \;\;
$[(T a)_\la b] = -\la [a_\la b], \quad
[a_\la (T b)] = (T+\la) [a_\la b]$,

\emph{skewsymmetry}  \;\;\,
$[a_\la b] = -[b_{-T-\la} a]$,

\emph{Jacobi identity}  \;\;
$[a_\la [b_\mu c]] -  [b_\mu [a_\la c]] = [[a_\la b]_{\la+\mu} c]$. 

\smallskip\noindent
The relationship between Lie conformal algebras and vertex algebras
is somewhat similar to the one between Lie algebras and their
universal enveloping associative algebras (see \cite{ref2,ref5}).
In \cite{ref5} we gave a definition of the notion of a vertex
algebra as a Lie conformal algebra equipped with one additional
product, which becomes the Wick product.
Lie conformal (super)algebras were classified in \cite{ref6};
their representation theory and cohomology theory 
were developed in \cite{ref7}.

\section{Lie pseudoalgebras}\label{s2}

\emph{Lie pseudoalgebras} are ``multi-dimensional'' 
generalizations of Lie conformal algebras \cite{ref8}. 
In the definition from the previous section, we consider
$\la$, $\mu$, etc., as $D$-dimensional vector variables,
we replace $T$ by $\bt=(T_1,\dots,T_D)$, and $\CC[T]$ by 
$\CC[\bt] \equiv \CC[T_1,\dots,T_D]$. 
For $D=0$ we let $\CC[\bt] \equiv \CC$; then a Lie pseudoalgebra
is just a usual Lie algebra.

\smallskip\noindent
\textbf{Examples} of Lie pseudoalgebras:
\\
 1. 
$\Cur \, \g = \CC[\bt] \tt\g$, \;
$[a_\bla b] = [a,b]$ for $a,b \in \g$, where $\g$ is a Lie algebra.
\\
 2. 
$W(D) = \CC[\bt]L^1 \oplus\cdots\oplus \CC[\bt]L^D$, \quad
$[{L^\al}_\bla L^\be] = (T_\al + \la_\al) L^\be + \la_\be L^\al$.
\\
 3. 
$S(D,\bchi) = \{ \, \sum P_\al(\bt) L^\al \st 
\sum \, (\d_{T_\al}+\chi_\al) P_\al(\T) = 0 \} \subset W(D)$, \quad
$\bchi \in \CC^D$.
\\
 4. 
$H(D) = \CC[\bt]L$, \; $D$ -- even, \;
$[L_\bla L] = \sum_{\al=1}^{D/2}
\bigl( \la_\al \, T_{\al+\frac{D}2} - \la_{\al+\frac{D}2} \, T_\al \bigr) L$.

\smallskip\noindent
More generally than in Example 1, a \emph{current pseudoalgebra} over a
Lie pseudoalgebra $R$ is defined by tensoring $R$ with
$\CC[T_1,\dots,T_{D'}]$ over $\CC[\T]$ and keeping the same $\la$-bracket
for elements of $R$, where $D'>D$.

\smallskip\noindent
\textbf{Theorem} (\cite{ref8})\textbf{.}
\emph{Every simple Lie pseudoalgebra, which is finitely generated over\/
$\CC[\bt]$, is a current pseudoalgebra over one of the above.}

\smallskip\noindent
In fact, in \cite{ref8} we introduced and studied a more general notion
of a Lie pseudoalgebra, in which $\CC[\bt]$ is replaced by the universal
enveloping algebra of a Lie algebra of symmetries.
Lie pseudoalgebras are closely related to the Lie--Cartan algebras
of  vector fields:
$W_D = \Der\, \CC[\![x_1,\dots,x_D]\!]$ (Witt algebra),
$S_D \subset W_D$ (divergence zero),
$H_D \subset W_D$ (hamiltonian), 
$K_D \subset W_D$ (contact). 
Lie pseudoalgebras are also related to Ritt's differential Lie algebras, 
linear Poisson brackets in the calculus of variations, classical 
Yang--Baxter equation, and Gelfand--Fuchs cohomology (see \cite{ref8}).
The irreducible representations of the above Lie pseudoalgebras
were classified in \cite{ref9}.

\section{Vertex algebras and vertex Lie algebras in higher dimensions}\label{s3}

``Multi-dimensional'' generalizations of vertex algebras were considered
in \cite{ref10,ref11}. The ones introduced by Nikolov in \cite{ref11}
arose naturally within a one-to-one correspondence with axiomatic 
\emph{quantum field theory} models satisfying the additional symmetry
condition of \emph{global conformal invariance} \cite{ref12}.
The theory of these vertex algebras was developed in \cite{ref11,ref13}.
The main difference with the usual vertex algebras discussed in Section 1
(which correspond to $D=1$)
is that now $\z=(z^1,\dots,z^D)$ is a vector variable. 
A \emph{field} on $V$ is defined as a linear map from 
$V$ to $V[\![\z]\!] [1/\z^2]$, where
$\z^2 \equiv \z\z = z^1 z^1 +\cdots+ z^D z^D$
(so the singularities of fields are supported on the light-cone).
Fields have a \emph{mode} expansion
\begin{equation*}
a(\z) = \sum_{n\in\ZZ} \, \sum_{m\in\ZZ_{\ge0}} \, \sum_{\si=1}^{\h_m} \,
a_{\{n,m,\si\}} \, (\z^2)^{n} \, h_{m,\si}(\z) 
, \qquad a_{\{n,m,\si\}} \in \End V,
\end{equation*}
where $\{ h_{m,\si}(\z) \}_{\si=1,\dots,\h_m}$
is a basis of the space of  harmonic 
homogeneous polynomials of degree $m$. 
Fields have the property
\begin{equation*}
(\z^2)^{N_{ab}} \, a(\z) b \in V[\![\z]\!] \equiv V[\![z^1,\dots,z^D]\!],
\qquad N_{ab} \in \NN,
\end{equation*}
and the \emph{locality} condition is now
\begin{equation*}
((\z-\w)^2)^{N_{ab}} \, [a(\z), b(\w)] = 0.
\end{equation*}
A vertex algebra $V$ is now endowed with $D$ commuting
\emph{translation operators\/} $T_1,\dots,T_D \in \End V$ satisfying
$[T_\al, a(z)] = \d_{z^\al} \, a(z) = (T_\al a)(z)$.
In \cite{ref13} we defined the \emph{residue} by 
$\Res \, a(\z) = a_{\{-\frac{D}2,0,1\}}$ if $h_{0,1}(\z) \equiv 1$; then
it is translation invariant and the modes of fields can be obtained as
residues. 
Introduce the notation $\iota_{\z,\w} F(\z-\w)$
for the formal Taylor expansion $\rme^{-\w \d_\z} F(\z)$.

\smallskip\noindent
\textbf{Theorem} (\cite{ref13})\textbf{.}
\emph{In any vertex algebra we have the} Borcherds identity 
\begin{equation*}
\begin{split}
a(& \z) b(\w) c \, \iota_{\z,\w} F(\z,\w)
-
b(\w) a(\z) c \, \iota_{\w,\z} F(\z,\w)
\\ 
& =
(\z^2)^{-L} \Bigl[ ( (\u+\z-\w)^2 )^{L} \, 
(\iota_{\z,\w} - \iota_{\w,\z}) \,
( a (\z-\w) b ) (\u) c \, F(\z,\w) 
\Bigr]_{\u=\w}
\end{split}
\end{equation*}
\emph{for $L\ge N_{ac}$ and}
$F(\z,\w) \in \CC[\z,\w,1/\z^2,1/\w^2,1/(\z-\w)^2]$.

\smallskip\noindent
For $F(\z,\w) = 1$ the above identity reduces to a \emph{commutator formula}.
Because of $\iota_{\z,\w} \!-\! \iota_{\w,\z}$ only the \emph{singular part} of 
$a (\z-\w)$ contributes, where
\begin{equation*}
a(\z)_\sip = \sum_{n\in\ZZ_{<0}} \, \sum_{m\in\ZZ_{\ge0}} \, 
\sum_{\si=1}^{\h_m} \,
a_{\{n,m,\si\}} \, (\z^2)^{n} \, h_{m,\si}(\z).
\end{equation*}
The singular parts of fields satisfy the \emph{Jacobi identity} \cite{ref13}
\begin{equation*}
\begin{split}
[& a(\z)_{\sip}, b(\w)_{\sip}] c
\\
& =
\Bigl(
\Bigl[ (\z^2)^{-L} ( (\u+\z-\w)^2 )^{L} 
(\iota_{\z,\w} - \iota_{\w,\z})
( a (\z-\w)_{\sip} b \bigr) (\u)_{\sip} c
\Bigr]_{\u=\w} \Bigr)_{\sip} .
\end{split}
\end{equation*} 
We also have \emph{translation invariance}
$[T_\al, a(\z)_\sip] = \d_{z^\al} \, a(\z)_\sip = (T_\al \, a)(\z)_\sip$
and \emph{skewsymmetry}
$a(\z)_\sip b = \bigl( e^{\z\T} (b(-\z)a) \bigr)_\sip$.
The above three axioms define the notion of a \emph{vertex Lie algebra}
in higher dimensions \cite{ref13}.
{}From a vertex Lie algebra one gets a vertex algebra by adding the
\emph{Wick product} $a_{\{0,0,1\}} b$, 
similarly to the construction of \cite{ref5} for $D=1$.

\smallskip\noindent
\textbf{Example.}
The modes of the \emph{real bilocal field\/} $V(\z,\w)$ 
from \cite{ref14}
obey the Lie algebra $\sp(\infty,\RR)$; hence it gives rise to a
vertex Lie algebra.
The conformal  Lie algebra $\so(D,2)$ can be embedded in a
suitably completed and centrally extended $\sp(\infty,\RR)$.
The corresponding vertex algebra is \emph{conformal} and \emph{unitary}.
Its unitary positive-energy representations were determined in~\cite{ref14}.


\section*{Acknowledgments}

I would like to express my deep gratitude to my teachers Emil Horozov and
Ivan Todorov from Sofia and Victor Kac from MIT. I am indebted to my 
collaborators A.~D'Andrea, A.~De Sole, A.~Kirillov Jr., N.~M.~Nikolov,
\linebreak K.-H.~Rehren, A.~A.~Voronov, and M.~Yakimov.
It is a pleasure to thank the organizers of the Colloquium, and especially
S.~Catto, for the marvelous scientific and cultural program.
Finally, I would like to thank my family and my wife Vesselina
for their love and support.


\end{document}